\RequirePackage{fix-cm}
\documentclass[twocolumn]{svjour3}          
\smartqed  
\usepackage{graphicx}
\usepackage{dcolumn}
\usepackage{bm}
\usepackage{subcaption}
\usepackage{hyperref}
\usepackage[utf8]{inputenc}

\journalname{Computing and Software for Big Science}
\begin{document}

\title{End-to-End Physics Event Classification with CMS Open Data}
\subtitle{Applying Image-Based Deep Learning to Detector Data 
for the Direct Classification of Collision Events at the LHC}

\author{M. Andrews, M. Paulini, S. Gleyzer, B. Poczos}
\institute{M. Andrews, M. Paulini \at
              Department of Physics \\
              Carnegie Mellon University\\
              Pittsburgh, USA\\
            \and
           S. Gleyzer \at
              Department of Physics \\
              University of Florida\\
              Gainesville, USA\\
            \and
           B. Poczos \at
              Machine Learning Department \\
              Carnegie Mellon University\\
              Pittsburgh, USA\\
}
\date{\today}

\maketitle

\begin{abstract}
This paper describes the construction of novel end-to-end image-based classifiers that directly leverage low-level simulated detector data to discriminate signal and background processes in proton-proton collision events at the Large Hadron Collider at CERN. To better understand what end-to-end classifiers are capable of learning from the data and to address a number of associated challenges, we distinguish the decay of the standard model Higgs boson into two photons from its leading background sources using high-fidelity simulated CMS Open Data. We demonstrate the ability of end-to-end classifiers to learn from the angular distribution of the photons recorded as electromagnetic showers, their intrinsic shapes, and the energy of their constituent hits, even when the underlying particles are not fully resolved, delivering a clear advantage in such cases over purely kinematics-based classifiers.
\keywords{end-to-end \and detector images \and machine learning \and deep learning \and CNN \and Resnet \and photon ID \and event classification \and mass sculpting \and LHC \and CMS \and open data \and higgs boson}
\end{abstract}

\section{\label{sec:Introduction}Introduction}

An important aspect of searches for physics beyond the standard model (SM) of particle physics at the CERN Large Hadron Collider (LHC) is the identification of signal events from their corresponding backgrounds. At the Compact Muon Solenoid (CMS) experiment~\cite{cms}, this task is accomplished by first reconstructing the low-level detector data into progressively more physically-motivated quantities \cite{pf} until arriving at tabular-like particle-level data. Traditional analysis approaches~\cite{higgs,daniel} use these condensed inputs to construct an event classifier that capitalizes on the decay structure or topology of the processes involved. While such approaches have been widely successful in understanding the SM, they potentially lose information in the process and limit more exhaustive searches for physics beyond the standard model (BSM).

In this paper, we propose a novel approach for particle physics event classification that directly utilizes maps of detector level deposits as inputs to a convolutional neural network (CNN) -- an \emph{end-to-end} event classification method. While others have developed conceptually similar approaches, these classifiers still rely on using reconstructed particle-level data to construct some pixelated (or other abstract) representation~\cite{jetimages,jetimages2,wahid,circles} of the detector. Such an approach introduces additional processing wherein information may potentially be lost. Our classification method differs fundamentally from previous approaches because we directly build the images from maps of the actual recorded particle showers or energy deposits in the detector (or their simulated equivalents) \emph{before} any particle processing is performed on the raw data. This means the event classifier gains direct access to the maximum possible information about the recorded event. One of the main technical challenges in extracting such information, however, is realizing coherent multi-subdetector images of the CMS experiment. This task is non-trivial given the irregular segmentation, granularity, and spacing of the detector. Therefore, the focus of this paper is developing an image construction strategy for handling the complexity of a detector like CMS in a way that best preserves the recorded information, and to validate the approach in a simplistic but physically reasonable event classification setting. We will deal mainly with the calorimeters and utilize a simplified representation of the tracking information, with a more detailed treatment of tracks left for future work.

While performing end-to-end event classification on full detector-view images potentially offers the opportunity to exploit event-level correlations to their fullest, the image construction techniques described here can equally be applied to create particle-view images for performing end-to-end particle identification first, as part of a more conventional ``factorized'' workflow, before being fed into a separate traditional event classifier. Some complications introduced by this factorized workflow, however, are how to order the particle-level information when feeding it to the aggregating event classifier, or whether a variable number of particles per event should be considered. The final event classification results can potentially be sensitive to how these ordering issues are addressed. For anything more than two particles, ordering becomes a non-trivial task requiring the introduction of an artificial ordering scheme~\cite{kyle,atlasbjetrnn,jetflavorFCN,toptagFCN,toptagRNN}, or an algorithm that is invariant with respect to the ordering or number of particles \cite{particlecloud,energyflow}. The fully end-to-end event classification approach avoids these issues altogether by relying solely on the natural spatial distribution of the particle showers in the fixed, full detector view. In this sense, the fully end-to-end event classification approach describes a general framework that can be applied to arbitrarily complex physics processes, as are found in some searches for physics beyond the standard model (BSM).
    
Although the question of pursuing a factorized versus an end-to-end approach to event classification merits further study, for the purposes of this paper, most of the issues discussed above are avoided by considering a physics process with only two outgoing particles. We therefore explore the decay of the SM Higgs boson to two photons versus its leading backgrounds using the 2012 CMS Simulated Open Data, in which a high-fidelity simulation of the CMS detector response is made available. Since we are working with a simple two-body diphoton decay, we can also introduce a classifier based purely on particle 4-momentum to illuminate the discussion of the end-to-end results. Note, this comparison is intended purely as a pedagogical reference to understand what the end-to-end classifier is able to learn. It is not indented to serve as a head-to-head benchmark comparison. A more appropriate benchmark test would supplement the 4-momentum classifier with some shower shape information as applied, for instance, in the CMS Higgs to diphoton analysis~\cite{higgs}.

This paper is organized as follows: in Section~\ref{sec:Samples} we introduce the data samples and event selection. Section~\ref{sec:CMSDetector} describes the CMS geometry. In Section~\ref{sec:Images} we discuss the detector image construction, while we outline our network and training procedure in Section~\ref{sec:NetworkTraining}. A review of our end-to-end electron vs. photon particle identification results are presented in Section~\ref{sec:ShowerClass}. The end-to-end classification of SM Higgs to diphoton events is then presented in Section~\ref{sec:EvtClass} and our conclusions are summarized in Section~\ref{sec:Conclusions}.

\section{\label{sec:Samples}Open Data Simulated Samples}
The 2012 CMS Open Data provide high-quality, simulated CMS data events that are utilized to evaluate the end-to-end approach. The CMS Open Data contains the highest grade of detector simulation available, using the \texttt{Geant4}~package~\cite{geant4} for a first-principles modeling of the interaction of particles with the CMS detector material, combined with the most detailed geometry model of the CMS detector available~\cite{fullsim}. It is noted that the studies in references~\cite{jetimages,jetimages2,wahid,kyle} used parametric-based simulations of the detector response with a simplified detector geometry~\cite{delphes}.

\paragraph*{Datasets.}
$\,$\\
For our signal sample, we choose the gluon fusion Higgs to diphoton dataset~\cite{h2gg}, $gg \rightarrow H \rightarrow \gamma\gamma$, with a Higgs mass of $m_H=125\,\mathrm{GeV}$. For the background samples, we choose two illustrative processes: quark fusion to prompt diphoton~\cite{born}, $q\bar{q} \rightarrow \gamma\gamma$, or the so-called Born diphoton production, and $\gamma+\mathrm{jet}$ production~\cite{gjet}. The $\gamma\gamma$ background is an irreducible background as it also contains two photons in the final state, differing only in their kinematics with photons from the $H \rightarrow \gamma\gamma$ decay. In the $\gamma+\mathrm{jet}$ background, the jet is electromagnetically enriched to deposit its energy primarily in the electromagnetic calorimeter via a neutral meson decaying to two merged photons. The jet thus appears as a single photon-like cluster. While there are other backgrounds involved in the Higgs to diphoton decay, the chosen backgrounds are representative of the most challenging types: kinematically-differentiated decays ($\gamma\gamma$) and particle shower-differentiated decays due to unresolved objects ($\gamma+\mathrm{jet}$). All the above processes were generated with the \texttt{Pythia\,6}~\cite{pythia6} package and account for the multi-parton interactions from the underlying event as well as pile-up (PU)~\cite{PU}. The PU distributions are run era dependent, ranging from a peak average PU of $\langle \mathrm{PU} \rangle =$~18-21.

\paragraph*{Event selection.}
$\,$\\
We categorize the samples based on pseudorapidity $\eta$, where $\eta=-\ln[\tan(\theta/2)]$ and $\theta$ is the spherical polar angle with respect to the beam axis. The \textbf{central} sample is restricted to $|\eta| < 1.44$ and the \textbf{central+forward} sample ranges up to $|\eta| < 2.3$, with the region around the electromagnetic calorimeter barrel-endcap boundary, $1.44 < |\eta| < 1.54$, excluded. For both categories, we require two reconstructed photons, each with transverse momentum $p_T > 20$~GeV. Since the number of events is limited and unbalanced between datasets, with the lowest number coming from the $\gamma\gamma$ dataset, we apply no further photon quality requirements. We require, however, that the reconstructed mass of the diphoton system is $m_{\gamma\gamma} > 90$~GeV. With these selections, we obtain 63,502 and 135,602 events in the $\gamma\gamma$ dataset for the central and central+forward categories, respectively. The selected events are broken down by run era in Table~\ref{table:Nevents}.

\begin{table*}[h]
\caption{Number of selected events by run era, per $|\eta|$ category, per dataset.}
\centering
\begin{tabular}{l c c c}
\hline\noalign{\smallskip}
Category & Run2012AB & Run2012C & Run2012D \\
\noalign{\smallskip}\hline\noalign{\smallskip}
Central         & 16308 & 24538 & 22206 \\
Central+forward & 35141 & 47885 & 52576 \\
\noalign{\smallskip}\hline
\end{tabular}
\label{table:Nevents}
\end{table*}

For the remaining datasets, we truncate them by taking the first $N_i$ events satisfying the same era $i$ breakdown, so that we only consider the same number of events per run era, per dataset, in order to minimize learning based on differences in pile-up or class proportion.

\section{\label{sec:CMSDetector}CMS Detector}
The Compact Muon Solenoid (CMS) detector is arranged as a series of concentric cylindrical sections---including a barrel section and circular endcap sections---that encloses a central interaction point where the LHC proton beams collide. Each cylindrical detector section or \emph{subdetector} specializes in measuring one or more aspect of the particles decaying from the collision. Together, the information from the different subdetectors is used to re-create as complete a picture as possible of the collision event, or \emph{event} for short.

\subsection{\label{subsec:CMSgeom}Geometry}
We focus on the three subdetectors most relevant for this study: the inner tracking system (Tracker), the electromagnetic calorimeter (ECAL), and the hadronic calorimeter (HCAL). The Tracker is the innermost cylindrical part of CMS and is responsible for detecting the hits associated with the tracks left by charged particles as they fly outward from the interaction point. This is reflected in the use of fine silicon segments that provide precise spatial resolution but no practical energy measurement. In 2012, the Tracker was composed of 13 barrel layers and 14 endcap layers. To avoid particles slipping through cracks in the layers, the barrel and endcap layers of the Tracker overlap in pseudorapidity in a non-trivial way. Each layer is composed of fine strip- or pixel-like silicon segments that provide the spatial localization. Moreover, the barrel and endcap sections of the Tracker are segmented differently: in cylindrical coordinates, with the beamline as the axis, they are in axis and azimuthal angle ($z,\phi$) in the barrel and in radius and azimuthal angle ($\rho,\phi$) in the endcap, with the dimensions of the segments changing with layer.

Surrounding the CMS Tracker system is the ECAL subdetector. The ECAL measures the energy deposits of electrons and photons by capturing almost all their energy using scintillating lead tungstate crystals. In the barrel section (EB), which spans $|\eta| < 1.479$, the ECAL is segmented by pseudorapidity ($i\eta_\mathrm{EB}$) and azimuthal angle ($i\phi_\mathrm{EB}$) giving a $170 \times 360$ crystal arrangement, respectively. This gives the EB an average granularity of $\Delta\eta_\mathrm{EB} \times \Delta\phi_\mathrm{EB} = 0.0174 \times 0.0174$. In the endcap sections (EE+/EE-), which span $ 1.479 < |\eta| < 3.0$, the crystals are arranged in a Cartesian grid ($iX,iY$) with 7324 crystals per endcap. For reference, most electrons/photons will deposit $>90\%$ of their energy within a $3 \times 3$ crystal window. The ECAL thus provides a potent combination of spatial and energy resolution.

Enclosing the ECAL is the HCAL subdetector that measures the energy deposits of hadronic particles---primarily the charged pions and kaons making up hadronic jets---using scintillating brass towers. In the barrel (HB) and endcap (HE+/HE-) sections, the HCAL is segmented in pseudorapidity ($i\eta_\mathrm{HCAL}$), azimuthal angle ($i\phi_\mathrm{HCAL}$), and readout depth ($d_\mathrm{HCAL}$). The depth segmentation varies with $|\eta|$ but is uniform in $\phi$. Combined, the HB and HE span the range $|\eta| < 3$, with the boundary between the two occurring at $|\eta| = 1.566$. The HCAL is considerably more coarse than ECAL at $\Delta\eta_\mathrm{HCAL} \times \Delta\phi_\mathrm{HCAL} = 0.087 \times 0.087$, or about $5 \times 5$ EB crystals per HB tower. Starting at $|i\eta_\mathrm{HCAL}| > 20$, the $\phi$ granularity in the HE becomes more coarse still with $\Delta\phi_\mathrm{HCAL} = 0.174$. Note that in the CMS Open Data, $i\phi_\mathrm{HCAL} = 1$ does \emph{not} correspond to the same plane as $i\phi_\mathrm{ECAL} = 1$, and thus must be shifted accordingly.

To avoid particles slipping through cracks undetected, none of the barrel-endcap boundaries between the Tracker, ECAL, and HCAL overlap.

\subsection{\label{subsec:reco}Reconstruction}
Below, we briefly describe how the particle interactions with the detector are used to reconstruct the detector hits which form the basis of the low-level data used in the end-to-end approach. For reference, we also provide an overview of how these low-level detector data are used to form the higher-level particle data conventionally used for physics analyses.

\paragraph*{Calorimeter Hit Reconstruction.}
$\,$\\
Since both the ECAL and HCAL are scintillating calorimeters, they share similar strategies to the energy reconstruction of calorimeter deposits or hits \cite{cms}. As an electromagnetic (hadronic) particle enters an ECAL crystal (HCAL tower), an electromagnetic (hadronic) shower is produced. This is detected as a light pulse which is digitized into a series of amplitude readings over time---amounting to a short video of the shower evolution in the ECAL crystal (HCAL tower). By fitting a pulse shape onto these digitized amplitudes, the energy and timing associated with this deposit can then be determined. These values are then calibrated to give a final reconstructed energy and time per ECAL crystal (HCAL tower), leading to what is known as the \emph{reconstructed hit}.

\paragraph*{Tracker Hit \& Track Reconstruction.}
$\,$\\
In contrast, as charged particles pass through the finely segmented Tracker subdetector, they deposit very little energy in the silicon. As such, the Tracker hits provide precise position information for charged particle track reconstruction but no practical energy information. Using the hits recorded in the different layers of the Tracker, a combinatorial Kalman-filter pattern recognition algorithm \cite{cms} is used to iteratively fit charged particle tracks through the Tracker hits starting from the seed layer. From these \emph{reconstructed track} fits, various track parameters can be obtained, in particular, the track's position at the point of closest approach to the beamline (perigee), and its transverse momentum from its bending in the magnetic field of the CMS solenoid.

\paragraph*{High-level Particle Reconstruction.}
$\,$\\
The reconstructed tracks and calorimeter hits are the basic inputs to the rule-based CMS Particle Flow algorithm \cite{pf} that constructs intermediate-level data before producing final, high-level particle data. These include attributes, such as probable particle identity, kinematics, and shower shape features. They serve as the primary inputs to most event classifiers used in CMS analyses. In contrast, in the end-to-end approach, the inputs are the reconstructed tracks and calorimeter hits. Due to the present unavailability of Tracker hits in the CMS Open Data, we use the reconstructed \emph{tracks} rather than the low-level \emph{hits}, similar to the approach in \cite{wahid}.

\section{\label{sec:Images}Detector Images}
The CMS Open Data contains information about the reconstructed hits for the ECAL (HCAL) subdetectors, making it possible to construct calorimeter images whose pixels correspond exactly to physical crystals (towers). This is important because not all crystals (towers) have the exact same dimensions and images created using averaged dimensions will incur some distortion. Such a level of accuracy would not be possible with intermediate-level data like calorimeter towers (which have an HCAL-like granularity) or the particle-level data which are no longer expressed in detector coordinates.

\paragraph*{Combining Images.}
$\,$\\
The main challenge in combining subdetector images arises not from differences in granularity but from differences in segmentation and the fact that regions of dissimilar segmentation overlap. For subdetector sections which do not spatially overlap (e.g. the ECAL barrel and the ECAL endcap) these images are kept separate. However, for subdetector sections which do overlap, such as the ECAL barrel and the HCAL barrel calorimeters, the depth information will be compromised if the images are not combined at the input level. Even though, in 2-dimensional CNNs, convolutions are not performed along the depth axis, the activations along the depth axis are still being summed over. 

\begin{figure*}[!htbp]
\begin{subfigure}{.65\textwidth}
  \includegraphics[width=\textwidth]{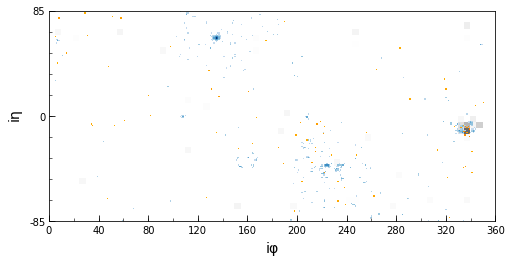}
  \caption{Barrel section of composite image in ECAL-centric geometry. Image resolution: 170 $\times$ 360.}
  \label{fig:EvtDisplay_EB}
\end{subfigure}%
\\
\begin{subfigure}{.8\textwidth}
  \includegraphics[width=.4\linewidth]{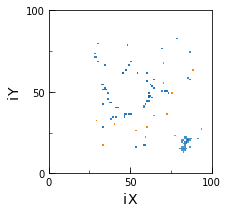}
  \includegraphics[width=.4\linewidth]{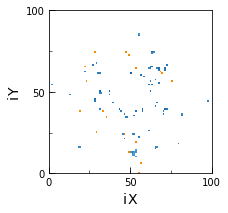}
  \caption{Endcap sections of composite image in ECAL-centric geometry. Image resolution: 100 $\times$ 100.}
  \label{fig:EvtDisplay_EE}
\end{subfigure}
\\
\caption{Composite images of a single $\gamma+\mathrm{jet}$ event in the ECAL-centric geometry with separate Barrel (\ref{fig:EvtDisplay_EB}) and Endcap (\ref{fig:EvtDisplay_EE}) images. Tracks are in yellow log scale, ECAL hits in blue log scale, and HCAL hits in gray linear scale. Additional zero suppression applied for clarity. Note the photon at around ($i\eta=70, i\phi=130$) which is free of HCAL hits or Tracks. In contrast, the jet at around ($i\eta=-10, i\phi=340$) shows contributions from all three subdetectors. Only the Barrel images (\ref{fig:EvtDisplay_EB}) are used for classification in the central category (see Section \ref{sec:Samples}).}
\label{fig:EvtDisplay_ECAL}
\end{figure*}

\begin{figure*}[!htbp]
\includegraphics[width=.65\textwidth]{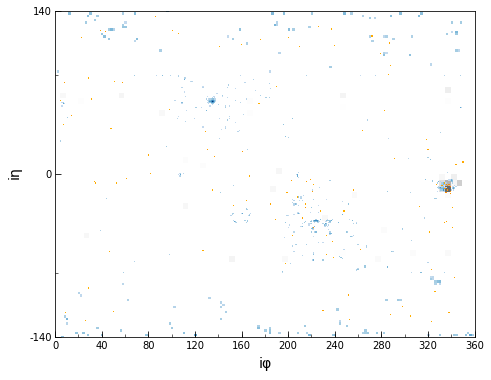}
\caption{Composite image of a single $\gamma+\mathrm{jet}$ event in the stitched Barrel+Endcaps HCAL-centric geometry. Extent of EB indicated by minor ticks on y-axis. Tracks are in yellow log scale, ECAL hits in blue log scale, and HCAL hits in gray linear scale. Additional zero suppression applied for clarity. See also description in Fig \ref{fig:EvtDisplay_ECAL}. Image resolution: 280 $\times$ 360. }
\label{fig:EvtDisplay_HCAL}
\end{figure*}

Therefore, to investigate the trade-off between detector fidelity and image integration, we experiment with different geometry strategies: we choose a subdetector $S$ to represent with the highest fidelity and project all other subdetectors $S'$ to the segmentation and boundaries of $S$. 
Procedures for constructing ECAL- and HCAL-centric geometries are described below and visualized for a single $\gamma+\mathrm{jet}$ event in Figures \ref{fig:EvtDisplay_ECAL} and \ref{fig:EvtDisplay_HCAL}.

\paragraph*{ECAL Images.}
$\,$\\
The ECAL image is defined by reconstructed hit energies and ECAL crystal coordinates. These are distinct for the EB and the EEs since they have different segmentation (see Section \ref{subsec:CMSgeom}). For the EB, we construct an unrolled rectangular $170 \times 360$ image. For the EEs, we inscribe each circular EE section in a square $100 \times 100$ image. These define the ECAL-centric geometry. Alternatively, for the HCAL-centric geometry, we construct a contiguous ECAL image by projecting the $(iX,iY)$-segmented EEs onto an EB-like ($i\eta,i\phi$) segmentation. These are then stitched to the ends of the EB image to form a single $280 \times 360$ image that spans the same $\eta$ range as the HCAL. Since this results in sparse showers in the endcap regions, we smear out each hit over a $2\times2$ window. While this is a potentially lossy transformation on the EEs, it will allow us to combine the full ECAL range with the HCAL into a single composite image.

\paragraph*{HCAL Images.}
$\,$\\
The HCAL image is defined in terms of reconstructed hit energies versus HCAL tower coordinates. These are shared by the HB and the HEs due to their similar segmentation. Since most events ($\approx 99\%$) in our samples only deposit in the first HCAL depth, we approximate the HCAL as a single depth by summing over $d_\mathrm{depth}$ for a given $(i\eta_\mathrm{HCAL}, i\phi_\mathrm{HCAL})$. In addition, some towers overlap in physical $\eta$ and are summed over as well to provide consistent alignment with the ECAL image. Above $|i\eta_\mathrm{HCAL}| > 20$, where the $\phi$ granularity is halved (see Section \ref{subsec:CMSgeom}), we share the energy across two $i\phi_\mathrm{HCAL}$ towers. We can thus construct a single, contiguous $56 \times 72$ image for the combined HB and HE. Without loss of information, this image is upsampled by a factor of 5 to produce a $280 \times 360$ HCAL image. This defines the HCAL-centric geometry. For the ECAL-centric geometry, the portions of this image which overlap with EB are left untouched while those which overlap with the EEs are detached and projected from their native ($i\eta,i\phi$) segmentation onto an EE-like $(iX,iY)$ segmentation, giving a $100 \times 100$ image per endcap.

\paragraph*{Tracker Images.}
$\,$\\
Because of the lack of Tracker hits in the CMS Open Data, the tracker image is constructed as a 2D histogram of the track reconstructed $(\eta,\phi)$ position at perigee in either the ECAL- or HCAL-centric geometry. To discriminate against the numerous pile-up tracks in an event, each track entry is weighted by its transverse momentum. Only \emph{high-purity} tracks~\cite{tkpurity}, or tracks with the highest level of fit quality, are used.

For the central category, we use only the subdetector images which overlap with the EB (Figure \ref{fig:EvtDisplay_EB}) giving image inputs of resolution 170 $\times$ 360. For the central+forward category, we use images which contain both the EB+EE (ECAL-centric: Figures \ref{fig:EvtDisplay_EB}+\ref{fig:EvtDisplay_EE} or HCAL-centric: Figure \ref{fig:EvtDisplay_HCAL}). These give image inputs of resolution 170 $\times$ 360 and 100 $\times$ 100 for the ECAL-centric geometry, and 280 $\times$ 360 for the HCAL-centric one. Lastly, while the event selection described in Section~\ref{sec:Samples} applies $\eta$ cuts on candidate photons, no such requirements are applied in the construction of the actual detector images in this paper, although this remains an option for future work.

\section{\label{sec:NetworkTraining}Network and Training}
At the heart of the end-to-end classifier is a CNN. In this Section, we describe how these deep learning networks are applied in order to extract information from the various subdetector images (see Section \ref{sec:Images}) in a way that best complements each subdetector's knowledge of the event. Afterwards, we discuss some of the challenges associated with using end-to-end classifiers, how we train them, and how we evaluate their performance in this study.

\subsection{\label{subsec:NetArch}Network Architecture}
\begin{figure}[!htbp]
\begin{subfigure}{.5\textwidth}
  \centering
  \includegraphics[width=.4\linewidth]{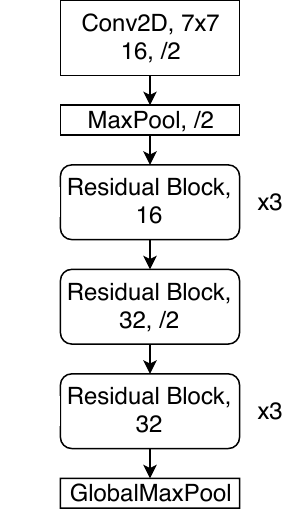}
  \caption{ResNet-15}
  \label{fig:ResNet_15}
\end{subfigure}
\\
\begin{subfigure}{.5\textwidth}
  \centering
  \includegraphics[height=.5\linewidth]{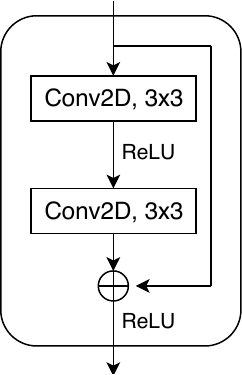}
  \caption{The Residual block with skip connection.}
  \label{fig:ResBlock}
\end{subfigure}%
\\
\begin{subfigure}{.5\textwidth}
  \centering
  \includegraphics[width=.8\linewidth]{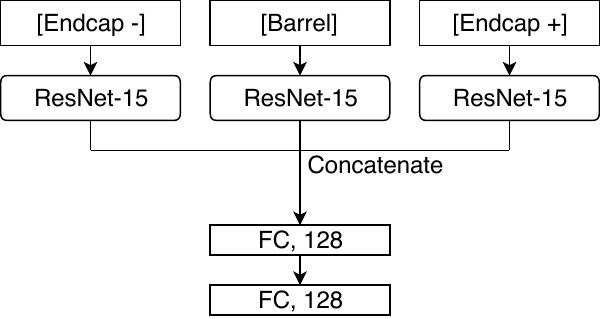}
  \caption{Concatenation of multiple ResNet-15 networks from separate barrel and endcap inputs.}
  \label{fig:ResNet_concat}
\end{subfigure}
\caption{The Residual Net (ResNet) architecture, as used for single (\ref{fig:ResNet_15}) and multiple (\ref{fig:ResNet_concat}) image inputs.}
\label{fig:ResNet}
\end{figure}

For all image-based classifiers, Residual Net-type networks (ResNet-15) are used due to their simplicity and scalability with image size and network depth \cite{DBLP}. A representative network is illustrated in Figure \ref{fig:ResNet_15}. Since image pixel intensities carry information about energy scale, the best performance is obtained when using MaxPooling operations with no batch normalization instead of AveragePooling. For samples in the central category, we use a single ResNet-15. For those in the central+forward category with ECAL-centric geometry, we use a separate ResNet-15 for each of barrel, endcap-, and endcap+. They are concatenated at the output of their final GlobalMaxPooling layer before being fed to a Fully-Connected Network (FCN), as illustrated in Figure \ref{fig:ResNet_concat}. In the central+forward HCAL-centric geometry, we use a single ResNet-15. The various end-to-end classifier models are summarized in Table \ref{table:Nets}.

\begin{table*}[h]
\caption{Summary of end-to-end models used in this paper. *NOTE: Models from the central category only use the barrel portion of the subdetector images (c.f. Figure \ref{fig:EvtDisplay_EB}).}
\centering
\begin{tabular}{l l l l l}
\hline\noalign{\smallskip}
Model & Category & Architecture & Inputs & No. of trainable parameters\\
\noalign{\smallskip}\hline\noalign{\smallskip}
EB     & Central     & ResNet-15          & ECAL*                & 88723\\
CMS-B  & Central     & ResNet-15          & Tracker, ECAL, HCAL* & 90291\\
ECAL   & Central+Fwd & 3 x ResNet-15, FCN & ECAL                 & 295187\\
CMS-I  & Central+Fwd & 3 x ResNet-15, FCN & Tracker, ECAL, HCAL  & 299891\\
CMS-II & Central+Fwd & ResNet-15          & Tracker, ECAL, HCAL  & 90291\\
\noalign{\smallskip}\hline
\end{tabular}
\label{table:Nets}
\end{table*}

Within the available statistics, the end-to-end results do not benefit from deeper networks or the inclusion of a FCN in the case of a single ResNet-15. Other variations on concatenating the networks for multiple images were attempted but were found to be less performant.

To serve as a reference for learning the event kinematics, we train a separate 3-layer, 128-node FCN on the reconstructed 4-momenta of the two candidate photons in each event, which we denote as the \textbf{4-momentum classifier}. Specifically, the networks are trained on the transverse momenta of the photons divided by the diphoton mass, $p_{T,i}/m_{\gamma\gamma}$, their pseudorapidities $\eta_i$, and the cosine of their azimuthal separation, $\cos(\phi_1 - \phi_2)$, where $i=1,2$ is the photon index, in order of decreasing $p_{T}$. The normalization by $m_{\gamma\gamma}$ ensures the classifier is uncorrelated with the mass of the Higgs boson~\cite{higgs}. Note, that this classifier serves as a purely kinematics-based reference and does not take into account information about the shape of the photon showers. The 4-momentum classifier results were not sensitive to the depth and width of the FCN network; it contains 50433 trainable parameters as configured for this study.

While the 4-momentum classifier has the advantage of having less model parameters to train, the end-to-end classifier must also learn to extract information from a higher-dimensional, more `raw' form of the data. Therefore, as with any CNN, it generally requires a larger training set for its potential to be fully realized, which may be an issue for physics samples which are computationally expensive to generate. As discussed in  Section~\ref{sec:EvtClass}, we do in fact see evidence that some of our results are constrained by training statistics.

\subsection{\label{subsec:Preproc}Preprocessing and Mass De-correlation}
Preprocessing plays a major role not just in improving the network optimization process but in controlling the physics content of the inputs themselves. In particular, for an event classifier intended for a resonance search, it is desirable for the classifier's output to \emph{not} be correlated with the mass of the signal resonance. Since one typically applies a cut on the classifier score to obtain a signal-enriched sample, this mitigates the risk of sculpting a false peak in the background.

Mass-sculpting---as it is commonly called---is especially an issue for irreducible backgrounds that differ only by kinematics and where good mass resolution is available from the classifier inputs. To measure it, we use the Cram\'{e}r-von Mises (CVM) metric suggested in \cite{CVM} and implemented in approximate form in \cite{CVMcode}. We calculate this on the classifier's signal score vs. reconstructed diphoton mass for true $\gamma\gamma$ events, as illustrated in Figure \ref{fig:MassCor_GG}.

To achieve mass de-correlation, we divide each image by the reconstructed diphoton mass for that event. To first approximation, this has the effect of transforming the energy scale of the diphoton system to have unit invariant mass for both signal and background events. Using the scalar sum of the diphoton $p_T$s (i.e. neglecting the angular component) is also effective. In fact, any quantity that maps the diphoton invariant mass for both signal and background events to a similar distribution achieves a similar effect. However, in the image-based approach, this only delays the onset of mass-sculpting but does not completely eliminate it---we suspect the photon shower profile provides an alternate avenue for learning the energy of the shower. In practice, while one can implement early stopping to intercept the training before the mass is learned, it is desirable to have a more robust and general method for de-correlating a classifier to a reconstructed variable like the mass. We accomplish this using an additional loss term proportional to the CVM metric itself, as outlined in Section~\ref{subsec:Train}.

Finally, because the energy deposits in the HCAL tend to be significantly lower in magnitude, we rescale the pixel intensities in the HCAL image by a constant to improve training.

\subsection{Training\label{subsec:Train}}
We train for three-class ($H\rightarrow\gamma\gamma$ vs. $\gamma\gamma$ vs. $\gamma+\mathrm{jet}$) classification with the normalized classification scores for each class label given by the softmax function. We use the ADAM adaptive learning rate optimizer \cite{adam} to minimize the cross-entropy loss. In general, we use an initial learning rate of $5\times 10^{-4}$, batch size of 360, and train for 60 epochs.

To maximize the limited statistics available in the CMS Open Data, the validation set doubles as our test set. A breakdown of the training and test set split is shown in Table \ref{table:TrainSplit}. For simplicity, both training and validation/test sets contain balanced proportions of the three classes. All training was done using the \texttt{PyTorch} \cite{pytorch} software library running on a single NVIDIA Titan X GPU.

To achieve more robust mass de-correlation, in addition to the cross-entropy loss, we minimize a loss term proportional to the CVM metric \cite{CVM} of the training batch. Specifically, we implement the $k$-Nearest Neighbors version of the CVM gradient with $k=12$ and a strength of $\lambda=0.15$ (central) or $0.2$ (central+forward). These values are tuned to give a CVM metric on the validation set of CVM $< 0.002$ for the central category and CVM $< 0.004$ for the central+forward category without the need for early stopping. While we have used the CVM loss in this context, in principle, this technique can be used to de-correlate the classifier from any other well-reconstructed variable or variables. A more detailed study of this matter will be treated in a separate paper.

\begin{table}[h]
\caption{Number of events in training and test sets \emph{for each class}. Test set doubles as validation set due to limited statistics. The total training and test sets contain a balanced proportion of class samples.}
\centering
\begin{tabular}{l c c}
\hline\noalign{\smallskip}
Category & Training Events & Test Events \\
 & per class & per class \\
\noalign{\smallskip}\hline\noalign{\smallskip}
Central         & 51200  & 11800 \\
Central+forward & 120000 & 15600 \\
\noalign{\smallskip}\hline
\end{tabular}
\label{table:TrainSplit}
\end{table}

Note that, in this paper, priority is given to presenting a broad and consistent survey of end-to-end classifiers over individual classifier optimization. As such, individual classifier hyper-parameter tuning was kept to a minimum, although we found the above training parameters to be robust across the different end-to-end classifiers.

\subsection{\label{subsec:Eval}Evaluation}
We use the area under the curve (AUC) of the normalized Receiver Operating Characteristic (ROC) as the main figure of merit in this paper but also present the false positive rate (FPR) at a fixed true positive rate (TPR) of 70\%. As is common in High Energy Physics, the ROC curve is interpreted in terms of the signal sample efficiency (TPR) vs. background sample rejection (1 - FPR). To evaluate the multi-class classification results, we define a per-class (1-vs-Rest) ROC and select the classifier with the best ROC AUC score in the signal label, subject to constraints on the CVM metric. To better understand the performance between the individual backgrounds in the multi-class classifier, for each background, we also present the (1-vs-1) signal vs. single background component of the ROC in the signal label. While the latter helps to give a sense of the individual background performance, it should be noted that multi-class classification is an inherently coupled problem, as is the nature of physics event classification. As outlined in Table \ref{table:TrainSplit}, the evaluation is done on a \emph{balanced} mix of class samples. While this is not necessarily the case in reality, it allows for a simple and unbiased assessment of classifier performance for several metrics.

Finally, to maximize the available training data in light of the limited statistics of the CMS Open Data, we apply the non-standard practice of using the validation set to present our final evaluation metrics. We expect any bias that this approach presents to be covered by our statistical uncertainties described below.

\subsection{\label{subsec:Syst}Uncertainties \& Optimality}
The dominant uncertainty in the evaluation metrics is due to statistical uncertainties associated with the finite validation set size (see Table \ref{table:Nevents}). We estimate this uncertainty to be of order $10^{-2}$ assuming Poisson statistics. While additional, systematic uncertainties may arise due to the mismodeling of the physics processes in the simulated data, these are beyond the scope of this work.

With respect to the optimality of the training procedure~\cite{MLuncert}, one additional source of variation in the evaluation metrics arises from the stochastic nature of the model training. To obtain an estimate of this effect for a given model and choice of hyper-parameters, one can vary the random seed used to feed the training data to the model. While this would be prohibitively computationally expensive to perform on all the end-to-end models considered in this study, we perform such studies for a few representative models over five variations of the random seed to derive a standard deviation for all of the metrics we present. Performing these variations for the \textbf{EB} model in the central category and for the \textbf{ECAL} model in the central+forward category (see Table~\ref{table:Nets}), we find this variation to be of order $10^{-3}$.

\section{\label{sec:ShowerClass}Shower Classification}

For reference, we recap our earlier results \cite{acat} in end-to-end classification of electromagnetic showers. This illustrates how the same image construction techniques described in Section \ref{sec:Images} can be used to do particle identification as part of a more conventional, factorized event classification workflow. Using just the ECAL barrel section, we successfully discriminated simulated electron- ($e^-$) vs. photon- ($\gamma$) induced showers. While not a practical task when track information is taken into account, when the ECAL information is taken in isolation, electron- and photon-induced showers appear nearly identical. Through higher-order effects such as bremsstrahlung due to the electron's interaction with the Tracker material, and bending from the CMS solenoid, that the electron shower becomes slightly smeared and asymmetric in $\phi$---an effect that is practically impossible to discern by eye. 

In Table \ref{table:ShowerClass}, we present the best-in-category results of using CNN-based (\textbf{CNN}), convolutional long short-term memory-based (\textbf{Conv-LSTM}) recurrent neural networks \cite{convlstm}, and fully-connected neural networks (\textbf{FCN}) on 32$\times$32 ECAL images centered on the shower maximum constructed out of various low-level data inputs. Our results suggest a preference for convolutional-based architectures, and that it is sufficient to use reconstructed hit energies (see Section \ref{subsec:reco}) to attain best results. Moreover, the range of these scores serves to illustrate how sensitive end-to-end classifiers are at processing low-level detector information even when the showers appear indistinguishable to the naked eye.

\begin{table}[h]
\caption{Best-in-Category results of $e^-$ vs. $\gamma$ shower classification on 32 $\times$ 32 ECAL Barrel images. \emph{Energy} inputs correspond to reconstructed hit energies, while \emph{digis} correspond to the series of digitized amplitudes vs. time (see Section \ref{subsec:reco}).}
\centering
\begin{tabular}{l l c}
\hline\noalign{\smallskip}
Category & Network, Input & ROC AUC \\
\noalign{\smallskip}\hline\noalign{\smallskip}
CNN  & VGG, energy       & 0.81 \\
LSTM & Conv-LSTM, digis  & 0.80 \\
FCN  & 3-layers, digis   & 0.77 \\
\noalign{\smallskip}\hline
\end{tabular}
\label{table:ShowerClass}
\end{table}

As a further step, we take the whole EB image instead of just the shower crop. The $e^-$ vs. $\gamma$ results are shown in Table \ref{table:ShowerClass2}. We observe a minimal loss in performance, owing to the CNN's ability to learn features in a translationally-invariant manner. More importantly, we see a marked improvement in the classification of uncorrelated particle gun pairs ($e^+e^-$ vs. $\gamma\gamma$) due to the CNN's having learned that the shower pairs must be either both electron-like or both photon-like. This suggests that topological complexity works in favor of the end-to-end approach and that the greatest gains may come from more challenging topologies.

\begin{table}[h]
\caption{Results of shower classification on full ECAL Barrel images.}
\centering
\begin{tabular}{l l c}
\hline\noalign{\smallskip}
Classification & Network, Input & ROC AUC \\
\noalign{\smallskip}\hline\noalign{\smallskip}
$e^-$ vs. $\gamma$          & ResNet, energy & 0.79\\
$e^+e^-$ vs. $\gamma\gamma$ & ResNet, energy & 1.00\\
\noalign{\smallskip}\hline
\end{tabular}
\label{table:ShowerClass2}
\end{table}

In sum, these results illustrate the ability of end-to-end classifiers to discern fine shower details in granular detectors like the CMS ECAL. However, when dealing with particles from real physics decays, there is the additional complexity introduced by \emph{kinematics}, which we turn our attention to next.

\section{\label{sec:EvtClass}Event Classification}

In a real physics process, energy and momentum conservation impose physical constraints on the allowed kinematics of the produced particles. The case of the Higgs boson decay and its related backgrounds is no exception. For the $\gamma\gamma$ background, the shower types are, in fact, identical to those of the $H\rightarrow\gamma\gamma$ decay and any differences are entirely due to kinematics. For the $\gamma+\mathrm{jet}$ background, in addition to kinematic differences, one of the particles is of a different type, and the opportunity exists to exploit differences in the particle shower shape, as we have already seen in the previous Section~\ref{sec:ShowerClass}. Moreover, in a realistic physics scenario, the classifier must also simultaneously discriminate between multiple decay processes. In this section, we therefore attempt to classify $H\rightarrow\gamma\gamma$ vs. $\gamma\gamma$ vs. $\gamma+\mathrm{jet}$ backgrounds.

The end-to-end event classification results are divided by pseudorapidity (see Section \ref{sec:Samples}), with the results for the central (central+forward) category given in Table \ref{table:MultiCentral} (Table \ref{table:MultiCentralFwd}), and corresponding ROC curves in Figure \ref{fig:MultiCentral} (Figure \ref{fig:MultiCentralFwd}). The ECAL-only classifier is labeled \textbf{EB} (\textbf{ECAL}) and the Tracks+ECAL+HCAL classifier in the ECAL-centric geometry is labeled \textbf{CMS-B} (\textbf{CMS-I}). For the central+forward region, we also include the results of the HCAL-centric classifier (\textbf{CMS-II}). In each category, we plot the signal vs. combined background ROC (1-vs-Rest), as well as the signal vs. single background ROC component (1-vs-1) (see Section \ref{subsec:Eval}). For context, we also include the results of the (mass de-correlated) 4-momentum-only classifier (\textbf{4-mom}) and the mass-aware ECAL-only classifier (\textbf{EB/ECAL, mass-aware}). Note that the results represent evaluations on balanced class samples (see Section~\ref{subsec:Eval}). Due to the limited training statistics available---which gives an edge to the 4-momentum classifier---the following results should not be taken as indications of ultimate end-to-end classifier performance.

\subsection{\label{subsec:Central} Central Pseudorapidity Region}

\begin{table*}[!htbp]
\caption{Multi-class Event Classification Results, central $|\eta| <1.44$ region. Uncertainties are on the last digit.}
\centering
\begin{tabular}{r c c c c}
\hline\noalign{\smallskip}
Metric & 4-mom & EB, mass-aware & EB & CMS-B \\
\noalign{\smallskip}\hline\noalign{\smallskip}
\textbf{1-vs-Rest:} ROC AUC / FPR@TPR=0.7\\
$H\rightarrow\gamma\gamma$
    & 0.71/0.41 & 0.93/0.08 & 0.80/0.27 & 0.81/0.26 \\
$\gamma\gamma$
    & 0.81/0.25 & 0.92/0.06 & 0.83/0.24 & 0.84/0.22 \\
$\gamma+\mathrm{jet}$
    & 0.81/0.22 & 0.95/0.01 & 0.94/0.02 & 0.96/0.02 \\
\textbf{1-vs-1:} ROC AUC / FPR@TPR=0.7\\
$H\rightarrow\gamma\gamma$ vs $\gamma\gamma$
    & 0.77/0.32 & 0.91/0.11 & 0.72/0.40 & 0.72/0.40 \\
$H\rightarrow\gamma\gamma$ vs $\gamma+\mathrm{jet}$
    & 0.78/0.28 & 0.97/0.02 & 0.94/0.07 & 0.96/0.04 \\
\textbf{CVM} & 0.002 & 0.080 & 0.002 & 0.002 \\
\noalign{\smallskip}\hline
\end{tabular}
\label{table:MultiCentral}
\end{table*}

\begin{figure}
\begin{subfigure}{.5\textwidth}
  \centering
  \includegraphics[width=.8\linewidth]{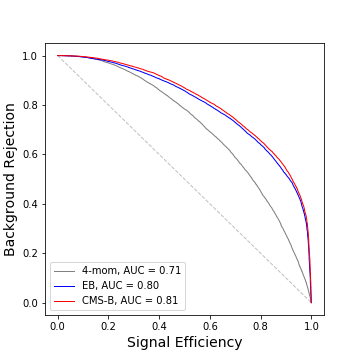}
  \caption{$H\rightarrow\gamma\gamma$ vs. Rest}
  \label{fig:Central_Multiclass_AUC_SgVAll}
\end{subfigure}%
\\
\begin{subfigure}{.5\textwidth}
  \centering
  \includegraphics[width=.8\linewidth]{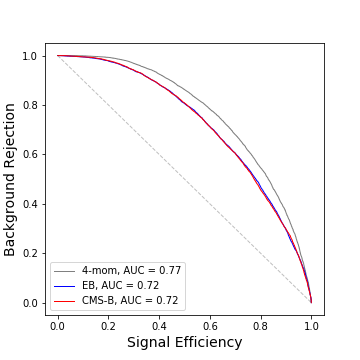}
  \caption{$H\rightarrow\gamma\gamma$ vs. $\gamma\gamma$ component.}
  \label{fig:Central_Multiclass_AUC_SgVGG}
\end{subfigure}
\\
\begin{subfigure}{.5\textwidth}
  \centering
  \includegraphics[width=.8\linewidth]{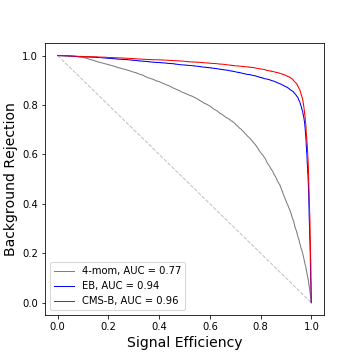}
  \caption{$H\rightarrow\gamma\gamma$ vs. $\gamma+\mathrm{jet}$ component.}
  \label{fig:Central_Multiclass_AUC_SgVGJ}
\end{subfigure}
\caption{Multi-class Event Classification ROC curves, central $|\eta|<1.44$ region.}
\label{fig:MultiCentral}
\end{figure}

To interpret these results, we first focus on the central category where we only use detector images from the barrel section of CMS. From the 1-vs-Rest plot (Figure \ref{fig:Central_Multiclass_AUC_SgVAll}), we see that, overall, image-based classifiers deliver substantially better performance versus purely kinematics-based classifiers. This is expected in the presence of a shower-differentiated background but serves to confirm that the end-to-end classifier is performing as expected. We also note that the \textbf{EB} and \textbf{CMS-B} classifiers perform comparably, with only negligible advantage to including additional subdetectors. Since the $\gamma\gamma$ background is a signature exclusively reconstructed in the ECAL, and the $\gamma+\mathrm{jet}$ background deposits the majority of its energy in the ECAL, this is expected. However, looking back at the event image in Figure \ref{fig:EvtDisplay_ECAL}, we see that other subdetectors carry significant noise from collision event pile-up and underlying event. It is thus not \emph{a priori} evident that the \textbf{CMS-B} classifier should perform as well as the \textbf{EB} under these circumstances. That no sizable degradation in performance is seen from the inclusion of additional noisy subdetectors indicates the ability of the end-to-end classifier to effectively screen out irrelevant features from extraneous image features. Lastly, we note that the highest performance---by a substantial margin---is achieved when the end-to-end classifier is allowed to learn the Higgs boson mass, or is \emph{mass-aware} (see Table \ref{table:MultiCentral}). To gain further insight into this behaviour, we next turn to the individual background components.

Looking at the binary $H\rightarrow\gamma\gamma$ vs. $\gamma\gamma$ component (Figure \ref{fig:Central_Multiclass_AUC_SgVGG}), the end-to-end classifier seems to under-perform compared to the \textbf{4-mom} classifier. We attribute this to the limited statistics available in training the end-to-end classifier. An indication of this statistical limitation is described in the following subsection for the Central+Forward category which contains more training statistics. This demonstrates that, at least for this study, we have paid no penalty in using a \emph{general} classifier trained on low-level detector data over a \emph{specialized} kinematics-based classifier that relied on our ability to reconstruct the event. The similarity in performance to the (mass de-correlated) 4-momentum classifier suggests the kinematic information is manifested in the detector image in two ways: the angular distribution of the photon showers and the absolute energy of the photon shower's constituent hits. While mass de-correlation is clearly a lossy operation, it preserves the angular information even though it removes the absolute energy scale, allowing for residual discrimination against irreducible backgrounds.

\begin{figure*}
\begin{subfigure}{\textwidth}
  \centering
  \includegraphics[width=.4\linewidth]{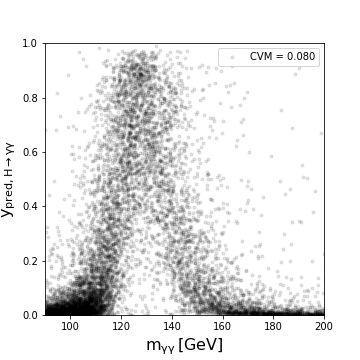}
  \includegraphics[width=.4\linewidth]{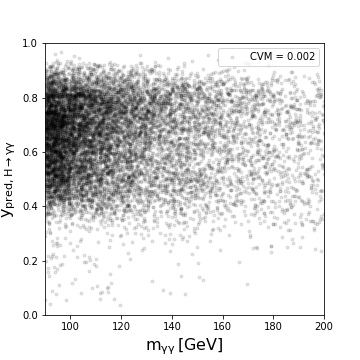}
  \caption{True $\gamma\gamma$ events, without mass de-correlation (left) and with (right).}
  \label{fig:MassCor_GG}
\end{subfigure}%
\\
\begin{subfigure}{\textwidth}
  \centering
  \includegraphics[width=.4\linewidth]{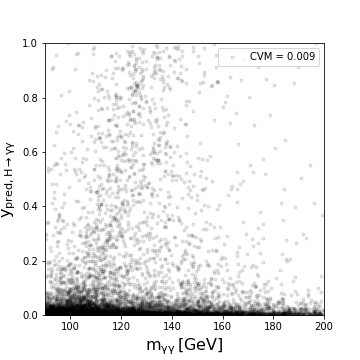}
  \includegraphics[width=.4\linewidth]{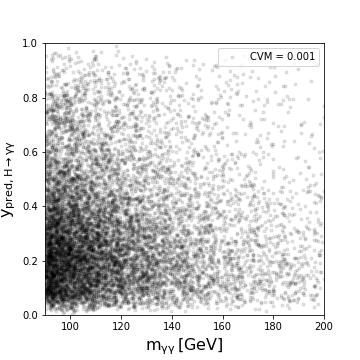}
  \caption{True $\gamma+\mathrm{jet}$ events, without mass de-correlation (left) and with (right).}
  \label{fig:MassCor_GJ}
\end{subfigure}%
\caption{Central \textbf{CMS-B} classifier output in signal label vs. reconstructed diphoton mass for true $\gamma\gamma$ (\ref{fig:MassCor_GG}) and $\gamma+\mathrm{jet}$ (\ref{fig:MassCor_GJ}) events, with and without mass de-correlation. The impact of de-correlation is more severe if the background lacks shower differentiation.}
\label{fig:MassCor}
\end{figure*}

Turning now to the $H\rightarrow\gamma\gamma$ vs. $\gamma+\mathrm{jet}$ component (Figure \ref{fig:Central_Multiclass_AUC_SgVGJ}), we see that this particular background is primarily responsible for the end-to-end advantage over the kinematics-only approach. This is expected because the jet (typically a merged $\pi^0\rightarrow\gamma\gamma$) is not fully resolved and is instead reconstructed as a single photon in the 4-momentum case, which is not supplemented with additional shower shape information. Despite registering as a single photon-like cluster, the jet appears in the ECAL image as a \emph{differentiated} shower, which, on occasion, is discernible by eye (see Figure \ref{fig:EvtDisplay_ECAL}). As reviewed in Section~\ref{sec:ShowerClass}, end-to-end classifiers are highly sensitive to differences in particle shower shapes even when no distinguishing kinematic information is present. Moreover, the $\gamma+\mathrm{jet}$ decay exhibits similar non-resonant kinematics to the $\gamma\gamma$ background and thus, to the 4-momentum classifier, the two should look alike. This is confirmed by their similar 4-momentum results (c.f. Figure \ref{fig:Central_Multiclass_AUC_SgVGG} vs. \ref{fig:Central_Multiclass_AUC_SgVGJ}). Owing to strong shower differentiation, the effect of mass de-correlation on the $\gamma+\mathrm{jet}$ background is much reduced. The impact of mass de-correlation depends strongly on the importance of kinematics---in particular, of the energy scale---against shower differentiation. For decays predominantly differentiated by kinematics, the effect of de-correlation will be substantial, while for primarily shower-differentiated decays, the effect will be minimal. In Figure~\ref{fig:MassCor}, we plot the classifier output in the signal label vs. diphoton mass for the two backgrounds to illustrate the impact of mass de-correlation.

\subsection{\label{subsec:CentralFwd}Central and Forward Pseudorapidity Region}

\begin{table*}[!htbp]
\caption{Multi-class Event Classification Results, central+forward $|\eta| <2.3$ region. Uncertainties are on the last digit.}
\centering
\begin{tabular}{r c c c c c}
\hline\noalign{\smallskip}
Metric & 4-mom & ECAL, mass-aware & ECAL & CMS-I & CMS-II\\
\noalign{\smallskip}\hline\noalign{\smallskip}
\textbf{1-vs-Rest:} ROC AUC / FPR@TPR=0.7\\
$H\rightarrow\gamma\gamma$ 
    & 0.73/0.38 & 0.93/0.07 & 0.81/0.26 & 0.83/0.23 & 0.81/0.25 \\
$\gamma\gamma$
    & 0.82/0.24 & 0.92/0.07 & 0.83/0.23 & 0.85/0.21 & 0.83/0.24 \\
$\gamma+\mathrm{jet}$
    & 0.81/0.21 & 0.96/0.01 & 0.94/0.03 & 0.95/0.02 & 0.96/0.01 \\
\textbf{1-vs-1:} ROC AUC / FPR@TPR=0.7\\
$H\rightarrow\gamma\gamma$ vs $\gamma\gamma$
    & 0.78/0.29 & 0.92/0.10 & 0.75/0.36 & 0.77/0.33 & 0.73/0.39 \\
$H\rightarrow\gamma\gamma$ vs $\gamma+\mathrm{jet}$
    & 0.79/0.26 & 0.97/0.02 & 0.94/0.06 & 0.95/0.04 & 0.95/0.04 \\
\textbf{CVM} & 0.004 & 0.088 & 0.004 & 0.004 & 0.003 \\
\noalign{\smallskip}\hline
\end{tabular}
\label{table:MultiCentralFwd}
\end{table*}

\begin{figure}
\begin{subfigure}{.5\textwidth}
  \centering
  \includegraphics[width=.8\linewidth]{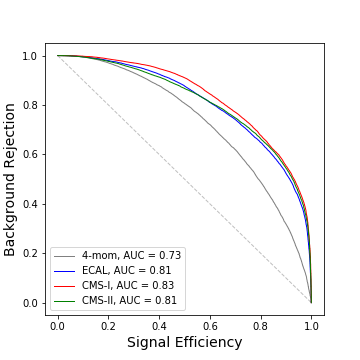}
  \caption{$H\rightarrow\gamma\gamma$ vs. Rest}
  \label{fig:CentralFwd_Multiclass_AUC_SgVAll}
\end{subfigure}%
\\
\begin{subfigure}{.5\textwidth}
  \centering
  \includegraphics[width=.8\linewidth]{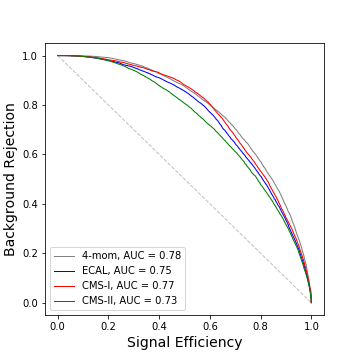}
  \caption{$H\rightarrow\gamma\gamma$ vs. $\gamma\gamma$ component.}
  \label{fig:CentralFwd_Multiclass_AUC_SgVGG}
\end{subfigure}
\\
\begin{subfigure}{.5\textwidth}
  \centering
  \includegraphics[width=.8\linewidth]{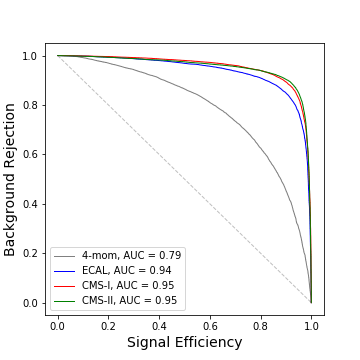}
  \caption{$H\rightarrow\gamma\gamma$ vs. $\gamma+\mathrm{jet}$ component.}
  \label{fig:CentralFwd_Multiclass_AUC_SgVGJ}
\end{subfigure}
\caption{Multi-class Event Classification ROC curves, central+forward $|\eta|<2.3$ region.}
\label{fig:MultiCentralFwd}
\end{figure}

In this category, we have included the Endcap images either in ECAL-centric (\textbf{ECAL}, \textbf{CMS-I}) or HCAL-centric (\textbf{CMS-II}) fashion (see Sections \ref{sec:Images} and \ref{subsec:NetArch}). In general, we find the qualitative conclusions from the central category to be also relevant for the central+ forward category. This informs us about the scalability of end-to-end network architectures and their ability to withstand the increased pile-up of the forward detector regions. The most notable difference is the significantly improved performance of the end-to-end classifiers in $H\rightarrow\gamma\gamma$ vs. $\gamma\gamma$ discrimination (c.f. Figure \ref{fig:Central_Multiclass_AUC_SgVGG} vs. \ref{fig:CentralFwd_Multiclass_AUC_SgVGG}) due to the larger available statistics for this category (see Section~\ref{subsec:Train}). This is to be contrasted with the \textbf{4-mom} classifier whose performance has mostly plateaued, and the $H\rightarrow\gamma\gamma$ vs. $\gamma+\mathrm{jet}$ discrimination (c.f. Figure \ref{fig:Central_Multiclass_AUC_SgVGJ} vs. \ref{fig:CentralFwd_Multiclass_AUC_SgVGJ}), which has mostly converged for both types of classifiers. This provides additional grounds to our claim that the end-to-end results for kinematic discrimination are statistically limited. In addition, we note that the ECAL-centric \textbf{ECAL, CMS-I} classifiers tend to outperform the HCAL-centric \textbf{CMS-II} classifier for $\gamma\gamma$ discrimination even though they have larger networks to train. This is, however, expected given the transformation applied to the ECAL endcaps (see Section~\ref{sec:Images}) and the role that spatial resolution plays in measuring particle kinematics for this background.

In sum, we find the biggest gains in discriminating backgrounds which have subtle shower differences as these maximally exploit both the full fidelity of the ECAL detector-level data and the CNN's ability to learn patterns at the level of individual image pixels.

\section{\label{sec:Conclusions}Conclusions}

In this paper, we described the construction of \emph{general}, end-to-end, image-based event classifiers, using high-fidelity simulated low-level CMS detector data as input. The use of end-to-end classifiers is not restricted to any particular topology and does not rely on the ability to fully reconstruct the event kinematics. It can be applied to arbitrarily complex event topologies and be particularly relevant to cases were traditional reconstruction approaches are difficult, for example, for highly boosted and merged topologies that arise in many BSM models \cite{h2aa}. To combine overlapping subdetector images of dissimilar segmentation, we chose one subdetector to render faithfully and projected all other subdetectors to its segmentation. While these classifiers are best suited to challenging decays, we have applied them in a simplified manner to the SM $H\rightarrow\gamma\gamma$ decay to highlight their key features and challenges. Through the irreducible $\gamma\gamma$ background, we were able to infer that such classifiers are able to learn about the angular distribution of the photon showers as well as the absolute energy of their constituent hits. We showed that we can definitively de-correlate the event classifier from the reconstructed diphoton mass while still preserving the angular information by using a CVM-based loss penalty. We found that such classifiers can learn about the photon shower shapes giving them an exceptionally strong advantage over a purely kinematics-based classifier in suppressing the reducible $\gamma+\mathrm{jet}$ background. Finally, we demonstrated the scalability and flexibility of the end-to-end classifiers when dealing with multiple detector images and networks, where they exhibited robustness against the presence of underlying event and pile-up.

\begin{acknowledgements}

We thank the entire CMS Collaboration for successfully recording LHC
proton-proton collision data as well as producing and releasing high
quality simulated data used in this paper. We also congratulate all
members in the CERN accelerator departments for the excellent
performance of the LHC and thank the technical and administrative staffs
at CERN and at other CMS institutes for their contributions to the
success of the CMS effort. In addition, we gratefully acknowledge the
computing centres and personnel of the Worldwide LHC Computing Grid for
delivering so effectively the computing infrastructure essential to CMS
analyses. Finally, we acknowledge the enduring support for the
construction and operation of the LHC and the CMS detector.

We would like to thank the CERN Open Data group for releasing their
simulated data under an open access policy. We strongly support
initiatives to provide such high-quality simulated datasets that can
encourage the development of novel but also realistic algorithms,
especially in the area of machine learning. We believe their continued
availability will be of great benefit to the high energy physics
community in the long run. Finally, M.A.~and M.P.~are supported by the
Office of High Energy Physics of the U.S.~Department of Energy (DOE)
under award DE-SC0010118. 

On behalf of all authors, the corresponding author states that there is no conflict of interest.
\end{acknowledgements}

\bibliographystyle{spphys}
\bibliography{biblio}{}

\end{document}